\documentclass[preprintnumbers,amsmath,amssymb,showpacs]{revtex4}

\usepackage{graphicx}
\usepackage{dcolumn}
\usepackage{bm}
\usepackage{booktabs}
\usepackage{CJK}

\begin{document}

\title{Exploration of photon-number entangled states using weak nonlinearities}

\author{Yingqiu He $^1$, Dong Ding $^{1,2}$, Fengli Yan $^{1}$}
\email{flyan@hebtu.edu.cn}
\author{Ting Gao $^3$}
\email{gaoting@hebtu.edu.cn}
\affiliation {$^1$ College of Physics Science and Information Engineering, Hebei Normal University, Shijiazhuang 050024, China \\
$^2$Department of Basic Curriculum, North China Institute of Science and Technology, Beijing 101601, China \\
$^3$College of Mathematics and Information Science, Hebei Normal University, Shijiazhuang 050024, China}
\date{\today}

\begin{abstract}
A method for exploring photon-number entangled states with weak nonlinearities is described. We show that it is possible to create and detect such entanglement at various scales, ranging from microscopic to macroscopic systems. In the present architecture, we suggest that the maximal phase shift induced in the process of interaction between photons is proportional to photon numbers. Also, in the absence of decoherence we analyze maximum error probability and show its feasibility with current technology.
\end{abstract}

\maketitle

\section{Introduction}
Motivated by the Knill-Laflamme-Milburn scheme for scalable quantum computing with linear optics \cite{KLM2001},  in recent years the field of quantum interference and entanglement with photonic qubits has grown rapidly \cite{Pan2012}.
Small numbers of entangled photons have been generated with stimulated parametric down-conversion \cite{PDC1995,SB2003} and used for implementing various quantum information protocols \cite{Kok2007}. More recently, several considerable methods for revealing macroscopic entangled states are reported with linear optics, such as by combining a single photon and a coherent beam on a beam splitter \cite{Macro-Entanglement2012}, amplifying and deamplifying a two-mode single photon entangled state \cite{GLS2013}, and so on.
For quantum entanglement of a large number of photons \cite{LHB2001, QE-LNumberP2004, MSV2008, SGS2014}, however, whether its fundamental principle or experimental demonstration is still a difficult and subtle task.

A cross-Kerr nonlinear medium is capable of inducing an interaction between the photons \cite{Imoto1985, MNS2005, Nonlinear-Interaction2014}, although its strength is very small for all experiments reported to date \cite{SI1996, LI2000, Optical-Fiber-Kerr2009}. Based on these weak nonlinearities one can implement photon-number quantum nondemolition measurement \cite{MNBS2005}, entanglement detection \cite{Barrett2005, ShengDengLong2010, DY2013}, quantum logic gates \cite{NM2004, Kok2008}, and miltiphoton entanglement \cite{DYG2013, KF2013, Micro-Macro-Entanglement, HDYG2015}. Since the nonlinearities are extremely weak, it seems natural to improve experimental methods so as to produce large enough nonlinear phase shifts and then follow the previous schemes without bound in the limit of weak nonlinearities. On the other hand, with current technology, it is also important to explore quantum circuit in the regime of weak nonlinearities  for  quantum information processing  \cite{LinHeBR2009, DY-PLA2013, DYG2014, YGC2011,GYE2014}.

In this Letter, we focus on the exploration of photon-number entangled states using weak nonlinearities. For each photon number $n$, we show a quantum circuit to evolve two-mode signal photons, ranging from microscopic to macroscopic systems (i.e., from $n=2$ to $n\gg 1$). In the regime of weak nonlinearities, more importantly, we consider the maximal phase shift induced in the process of interaction between photons satisfying $n \theta \simeq 10^{-2}$.
Moreover, in the absence of decoherence, we analyze error probability caused by the final homodyne measurement.

\section{Exploration of photon-number entangled states}

In Fock space, consider an arbitrary two-mode $n$-photon-number state
\begin{eqnarray}\label{}
|\Psi_n\rangle& &:= \sum_{l=0}^{[n/2]}|\psi_n^l \rangle_{s_1s_2}  =\sum_{l=0}^{[n/2]}(a_{l}|n-l,l\rangle_{s_1s_2}+b_{l}|l,n-l\rangle_{s_1s_2}).
\end{eqnarray}
Here
\begin{equation}\label{}
|\psi_n^l \rangle_{s_1s_2}=a_{l}|n-l,l\rangle_{s_1s_2}+b_{l}|l,n-l\rangle_{s_1s_2}, l=0,1,2,\cdots,[{n/2}]
\end{equation}
are a class of photon-number entangled states, two positions in the ket indicate respective the number of photons in two spatial modes $s_1$ and $s_2$, and $a_{l}$ and $b_{l}$ are complex parameters satisfying the normalization condition $\sum_{l} (|a_{l}|^2+|b_{l}|^2) = 1$. Obviously, $n$-photon number state (1) includes some canonical entangled number states \cite{NOON2000}, such as single-photon entangled state and the NOON state.  Theoretically, state (1) can be conditionally produced by letting $n$ photons pass through a beam splitter and the two spatial modes $s_1$, $s_2$ correspond to two outputs of beam splitter. Throughout the subsequent context, let $l=[n/2]-m$ with $m=0,1,2,\cdots,[n/2]$ and we replace $|\psi_n^l\rangle_{s_1s_2}$ by $|\psi_n^m\rangle_{s_1s_2}$ for simplicity.
For a given photon number $n$, we next show a method to detect the states $|\psi_n^m\rangle_{s_1s_2}$ by using weak nonlinearities.

Suppose there exist $n$ photons traveling through two spatial modes $s_1$ and $s_2$, namely signal modes; and we introduce a coherent state
$|\alpha\rangle=\text{exp}(-\frac{1}{2}|\alpha|^{2})\sum^{\infty}_{n=0}\frac{\alpha^{n}}{\sqrt{n!}}|n\rangle$ in probe mode; let $\theta$ and $n \theta$ be respective phase shifts on the coherent probe beam according to the two signal modes, as shown in Fig.1. In order to avoid inducing $-\theta$ in the interacting process we herein introduce a single phase gate, i.e., $R_n(\theta)=-\frac{1}{2}n(n+1)\theta$.
After an $X$ homodyne measurement on the probe beam plus appropriate local phase shift operation ${\phi_m}\left( x \right)$ on one of the signal modes using classical feed-forward information, at last, the original state can be projected into one of the photon-number entangled states $|\psi_n^m\rangle_{s_1s_2}$.

\begin{figure}
  \includegraphics[width=3.5in]{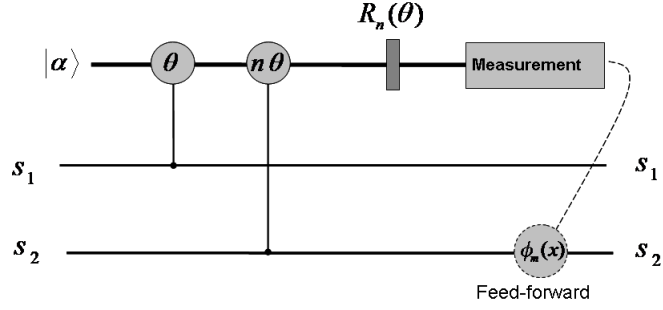}\\
  \caption{The schematic diagram of exploration of photon-number entangled states using weak nonlinearities. Consider $n$ photons traveling through two spatial modes $s_1$ and $s_2$ (say signal modes). $|\alpha\rangle$ is a coherent state in probe mode. $\theta$ and $n \theta$ are respective phase shifts on the coherent probe beam with several weak cross-Kerr nonlinearities. $R_n(\theta)$ is a single phase gate used to evolve coherent state. ${\phi_m}\left( x \right)$ represents a phase shift on one of the signal modes based on the classical feed-forward information.}
  \label{}
\end{figure}

We now describe our method in  details. For $n$ is even, after the interaction between the photons and the action of the phase gate, the combined system $|\Psi_n\rangle\otimes|\alpha\rangle\equiv|\Psi_n\rangle|\alpha\rangle$ evolves as
\begin{equation}\label{}
\sum_{m=0}^{n/2}[a_{m}|n/2+m, n/2-m\rangle_{s_1s_2}|\alpha\texttt{e}^{-m(n-1)\texttt{i}\theta}\rangle
+b_{m}|n/2-m, n/2+m\rangle_{s_1s_2}|\alpha\texttt{e}^{m(n-1)\texttt{i}\theta}\rangle].
\end{equation}
After the $X$ homodyne measurement on the probe beam \cite{Barrett2005, DYG2014}, the signal photons become
\begin{equation}\label{}
\sum_{m=0}^{n/2}f\{{x,\alpha \cos [m(n-1)\theta] }\}(a_{m}\texttt{e}^{-\texttt{i}\phi _{m}(x)} |n/2+m,n/2-m\rangle_{s_1s_2}
+b_{m}\texttt{e}^{\texttt{i}\phi _{m}(x)} |n/2-m,n/2+m\rangle_{s_1s_2}),
\end{equation}
where $f\left( {x,\beta } \right) = {\left( {2\pi } \right)^{ - 1/4}}{\texttt{e}^{ - \left( {x - 2{\beta}} \right)^2/4}}$ and ${\phi _m}\left( x \right) = \alpha \sin [m(n-1) \theta] \{ {x - 2\alpha \cos [m(n-1) \theta] } \} \bmod 2\pi$, $m=0,1, 2, \cdots, n/2$.
The functions $f \{ {x,\alpha \cos [m(n-1)\theta] } \}$, $m=0, 1, 2, \cdots, n/2$, are respective Gaussian curves with  peaks located at $2\alpha \cos [m(n-1)\theta]$ and these curves correspond to the probability amplitudes associated with the outputs of the signal photons.
${\phi_m}\left( x \right)$ are respective phase shift operations corresponding to the values of the $X$ homodyne measurement.
The midpoints between two neighboring peaks are designated as ${x_{{m_{k}}}} = \alpha \{ {\cos [(n/2-k-1)(n-1)\theta]  + \cos [(n/2-k)(n-1)\theta]} \}$, $k=0, 1, 2, \cdots, n/2-1$.
Note that with these $n/2$ midpoint values ${x_{{m_{k}}}}$, one can separate results of the homodyne measurement into $n/2+1$ intervals and then the input state can be projected into one of the states $|\psi_n^m\rangle_{s_1s_2}$ up to a phase shift operation $\phi_m(x)$ on one of the signal modes.
Clearly, for $x < {x_{{m_{0}}}}$ we observe immediately  the output state  $|\psi_n^{0}\rangle_{s_1s_2}$; for $x_{{m_{k-1}}} < x < x_{{m_{k}}}, k=1,2, \cdots, n/2-1$, we obtain the states $|\psi_n^{k}\rangle_{s_1s_2}$; and for $x > {x_{{m_{n/2-1}}}}$ we obtain the state $|\psi_n^{n/2}\rangle_{s_1s_2}$.
Considering there exist small overlaps between two neighboring curves, the error probabilities are thus given by $\varepsilon_k = { \textrm{erfc}\left( {{x_{d_k}}/2\sqrt 2 } \right)}/2$, where ${x_{{d_k}}} = 2\alpha \{{\cos [(n/2-k-1)(n-1)\theta]  - \cos [(n/2-k)(n-1)\theta] } \} \approx (n-2k-1)(n-1)^2\alpha {\theta ^2}$ are the distances of two nearby peaks.

Similarly, for $n$ is odd, the combined system then evolves as
\begin{equation}\label{}
\sum_{m=0}^{(n-1)/2}[a_{m}|(n+1)/2+m,(n-1)/2-m\rangle_{s_1s_2} |\alpha\texttt{e}^{-{\frac{1}{2}(2m+1)}(n-1)\texttt{i}\theta}\rangle
+ b_{m}|(n-1)/2-m,(n+1)/2+m\rangle_{s_1s_2}|\alpha\texttt{e}^{{\frac{1}{2}(2m+1)}(n-1)\texttt{i}\theta}\rangle].
\end{equation}
After the measurement on the probe beam, the signal photons become
\begin{eqnarray}\label{}
& & \sum_{m=0}^{{(n-1)}/2}f\{{x,\alpha \cos [{\frac{1}{2}(2m+1)}(n-1)\theta }]\} \nonumber \\
& & \times[a_{m}\texttt{e}^{-\texttt{i}\phi_{(2m+1)/2}(x)}|(n+1)/2+m,(n-1)/2-m\rangle_{s_1s_2} + b_{m}\texttt{e}^{\texttt{i}\phi _{(2m+1)/2}(x)}|(n-1)/2-m,(n+1)/2+m\rangle_{s_1s_2}],
\end{eqnarray}
where in order to simplify the notations we make use of the same expressions as the case of even $n$ for the functions $f\{{x,\alpha \cos [{\frac{1}{2}(2m+1)}(n-1)\theta }]\}$ and $\phi_{(2m+1)/2}(x), m=0,1,2,\cdots,{(n-1)/2}$. The peaks of Gaussian curves locate at $2\alpha \cos [\frac{1}{2}(2m+1)(n-1)\theta]$. Then, we can derive the same results of the midpoints ${x_{{m_{k}}}}$ between two neighboring peaks  and the distances ${x_{{d_k}}}$ separated by two nearby peaks.

By now, in the regime of weak Kerr nonlinearities we have described the evolution of $n$ photon combined system.
We next discuss some intriguing applications of the exploration of photon-number entangled states.

\section{Applications}
In many quantum measurements for linear optical quantum computation, one should always attempt to detect the signal photons and then project them into a desired subspace. Surprisingly, for a given photon number $n$, the setup in our scheme can be used for these processes based on postselection and classical feed-forward (an additional manipulation).
So, a direct application of the present scheme leads to an entangling gate for the states $\{|\psi_n^m\rangle_{s_1s_2}, m=0,1,2,\cdots,[{n/2}]\}$, and especially, for $a_m=b_m$ it yields one of the maximally entangled number states $(|n-m,m\rangle_{s_1s_2}+|m,n-m\rangle_{s_1s_2})/\sqrt{2}$. Moreover, it is easy to see that the present scheme is  suitable for constructing two-qubit polarization parity gate (see Fig.2 in Ref.\cite{NM2004}), discriminating between $|\psi_2^0\rangle$ and $|\psi_2^1\rangle$, nondestructively (see Fig.1 in Ref.\cite{Barrett2005}, also see Fig.2 in Ref.\cite{HDYG2015}), and so on.

Another important application of the present scheme is as an analyzer for the states $\{|\psi_n^m\rangle_{s_1s_2}\}$. Consider a state belonging to the set of states $\{|\psi_n^m\rangle_{s_1s_2}\}$ in signal modes. After the evolution of a series of optical devices followed by an
$X$ homodyne measurement on the probe beam, based on the value of the measurement one can infer immediately what the input must have been with a small error probability. Then, a conditional phase shift operation on one of the modes is necessary to restore the output state to that identified. In other words, the suggested analyzer of the states $\{|\psi_n^m\rangle_{s_1s_2}\}$ is nondestructive and thus the unconsumed signal photons can be recycled for further use.

\section{discussion and summary}

Note that the cross-Kerr nonlinearities are extremely weak and the order of magnitude of them is only $10^{-2}$ even by using electromagnetically induced transparency \cite{SI1996, LI2000}. In the present scheme,  let $n \theta = 1.0\times10^{-2}$ and then obtain $\varepsilon_{\text{max}} = { \textrm{erfc}\left[ {(1-1/n)^2 \alpha \times10^{-4} /2\sqrt2} \right]}/2$. Therefore, by applying an appropriate coherent probe beam the present scheme can has a small enough error probability and then be realized in a nearly deterministic manner.
Given $(1-1/n)^2 \alpha = 4\sqrt{2}\times10^{4}$ with $n=2,3,\cdots$, for example, then we have $\varepsilon_{\text{max}} \simeq 0.003$. Clearly, let $n=2$ and  mean photon numbers of coherent state $n_{\alpha}=|\alpha|^2=5.12\times10^{10}$, then the above value of error probability holds. Also, when $n\gg1$ and letting $n_{\alpha}= 3.2 \times 10^{9}$ we can also have the given error probability.

In summary, we show an architecture of exploration of photon-number entangled states using weak nonlinearities. Also, we suggest some interesting applications of the present scheme and analyze its error probabilities. The present scheme has two remarkable advantages. First, our scheme is feasible with the current experimental technology, because there is no large phase shift ($-\theta$ with $\theta\leq10^{-2}$, for example) in the interacting process with weak Kerr nonlinearities and then the strength of the nonlinearities we required are orders of magnitude in current practice. Second, by analyzing the error probability we show that our scheme works in a nearly deterministic way.

\section{Acknowledgements}
This work was supported by the National Natural Science Foundation of China under Grant Nos: 11475054, 11371005, Hebei Natural Science Foundation of China under Grant No:  A2014205060,  the Fundamental Research Funds for the Central Universities of Ministry of Education of China under Grant No:3142014068, Langfang Key Technology Research and Development Program of China under Grant No: 2014011002.

\end{document}